\documentclass[usenatbib,useAMS,usedcolumn]{mn2e}
\usepackage{longtable}
\usepackage{psfig}
\usepackage{epsfig}
\usepackage{graphicx}
\usepackage{times}
\usepackage{euscript}

\def\arcm{\hbox{$^\prime$}}
\def\etal{{\rm et al.}\thinspace}

\def\ie{{\it i.e.\ }}

\def\deg{\hbox{$^\circ$}}
\def\spose#1{\hbox to 0pt{#1\hss}}
\def\gtsim{$\mathrel{\spose{\lower 3pt\hbox{$\sim$}}
        \raise 2.0pt\hbox{$>$}}$\thinspace}
\def\simpropto{$\mathrel{\spose{\lower 3pt\hbox{$\sim$}}
        \raise 2.0pt\hbox{$\propto$}}$\thinspace}
\newcommand{\rosat}{\emph{ROSAT}}
\newcommand{\chandra}{\emph{Chandra}}

\newcommand{\asca}{\emph{ASCA}}

\newcommand{\arcs}{\mbox{\arcm\hskip -0.1em\arcm}}
\newcommand{\Lx}{\ensuremath{L_{\mathrm{X}}}}

\newcommand{\Zsol}{\ensuremath{Z_{\odot}}}

\newcommand{\Msol}{\ensuremath{M_{\odot}}}
\newcommand{\LB}{\ensuremath{L_{\mathrm{B}}}}
\newcommand{\LBsol}{\ensuremath{L_{B\odot}}}

\newcommand{\LxLbtwo}{\ensuremath{\mbox{\Lx :\LB}}}

\newcommand{\Bfit}{\ensuremath{\beta_{fit}}}

\newcommand{\NH}{\ensuremath{N_{\mathrm{H}}}}

\newcommand{\s}{\ensuremath{\mbox{~s}}}
\newcommand{\ps}{\ensuremath{\s^{-1}}}
\newcommand{\cm}{\ensuremath{\mbox{~cm}}}
\newcommand{\pcmsq}{\ensuremath{\cm^{-2}}}
\newcommand{\km}{\ensuremath{\mbox{~km}}}
\newcommand{\Mpc}{\ensuremath{\mbox{~Mpc}}}
\newcommand{\pMpc}{\ensuremath{\Mpc^{-1}}}
\newcommand{\kmpspMpc}{\ensuremath{\km \ps \pMpc\,}}
\newcommand{\erg}{\ensuremath{\mbox{~erg}}}
\newcommand{\ergps}{\ensuremath{\erg \ps}}
\newcommand{\ergpspcmsq}{\ensuremath{\erg \ps \pcmsq}}
\newcommand{\kmps}{\ensuremath{\km \ps}}

\newcommand{\Ho}{\ensuremath{H_\mathrm{0}}}

\newcommand{\ML}{\ensuremath{\mbox{\Msol/\LBsol}}}
\newcommand{\Rth}{\ensuremath{R_{\mathrm{200}}}}
\newcommand{\Dtf}{\ensuremath{D_{\mathrm{25}}}}

\begin{document}

\title[
The isolated elliptical NGC~4555 observed with Chandra
]
{
The isolated elliptical NGC~4555 observed with Chandra
}
\author[
E. O'Sullivan and T.~J. Ponman 
]
{
E. O'Sullivan\footnotemark$^{1}$, T.~J. Ponman$^{2}$\\
$^{1}$ Harvard Smithsonian Center for Astrophysics,
60 Garden Street, Cambridge, MA 02138, USA\\
$^{2}$ School of Physics and Astronomy, University of Birmingham,
Edgbaston, Birmingham B15 2TT\\
\\
}

\date{Accepted 2004 ?? Received 2004 ??; in original form 2004 ??}
\pagerange{\pageref{firstpage}--\pageref{lastpage}}
\def\LaTeX{L\kern-.36em\raise.3ex\hbox{a}\kern-.15em
    T\kern-.1667em\lower.7ex\hbox{E}\kern-.125emX}

\label{firstpage}

\maketitle

\begin{abstract}
  We present analysis of a \chandra\ observation of the elliptical galaxy
  NGC~4555. The galaxy lies in a very low density environment, either
  isolated from all galaxies of similar mass or on the outskirts of a
  group. Despite this, NGC~4555 has a large gaseous halo, extending to
  $\sim$60 kpc. We find the mean gas temperature to be $\sim$0.95 keV and
  the Iron abundance to be $\sim$0.5\Zsol. We model the surface brightness,
  temperature and abundance distribution of the halo and use these results
  to estimate parameters such as the entropy and cooling time of the gas,
  and the total gravitational mass of the galaxy. In contrast to recent
  results showing that moderate luminosity ellipticals contain relatively
  small quantities of dark matter, our results show that NGC~4555 has a
  massive dark halo and large mass-to-light ratio (56.8$^{+34.2}_{-35.8}$
  \ML\ at 50 kpc, 42.7$^{+14.6}_{-21.2}$ at 5$r_e$, 1$\sigma$ errors). We
  discuss this disparity and consider possible mechanisms by which galaxies
  might reduce their dark matter content.

\end{abstract}

\begin{keywords}
galaxies: individual: NGC 4555 -- X-rays: Galaxies -- galaxies: elliptical
and lenticular, cD
\end{keywords}

\footnotetext{Email: ejos@head.cfa.harvard.edu}

\section{Introduction}
\label{sec:intro}
The majority of galaxies are found in groups and clusters \citep{Tully87},
and this is particularly true of elliptical galaxies. The
morphology-density and morphology-radius relations show that elliptical
galaxies are most common in the cores of clusters and groups
\citep{Dressler80,MelnickSargent77}. In a hierarchical model of structure
formation, this can be explained as a product of the processes which form
ellipticals. The merger hypothesis \citep{Toomres72} suggests that the
product of the merger of two spiral galaxies will be an elliptical
galaxy. If this is the case, then galaxy groups are the most likely
location of elliptical formation, as these systems have high galaxy
densities but relatively low velocity dispersions. The merger of groups to
form larger galaxy clusters naturally leads to a large population of
elliptical galaxies in the most massive systems. 

Given this model of elliptical galaxy formation, it is perhaps unsurprising
that the best known elliptical galaxies are found in galaxy groups and
clusters. Most clusters and many groups are dominated by a giant elliptical
(or cD) galaxy which lies at the bottom of the cluster potential well.
X-ray observations have shown these objects to be surrounded by haloes of
highly luminous gas \citep{Formanjonestucker85,Trinchierietal86}. The
clusters and groups often have their own haloes of hot X-ray emitting gas
\citep{Kelloggetal75}, and models of galaxies in the cores of such systems
suggest that the galaxy halo is probably enhanced by inflow of gas from the
surrounding intra-cluster medium \citep{MathewsBrighenti03}. Because of
their high X-ray luminosities, these galaxies are the most easily observed
in X-rays, and most detailed analyses of elliptical galaxies focus on them
\citep[\textit{e.g.}][]{Buoteetal03a,Sakelliouetal02,Jonesetal02}.

However, the location of these galaxies in a dense environment, surrounded
by a reservoir of high temperature gas means that their intrinsic
properties must always be in doubt. In particular, the question of whether
elliptical galaxies produce the observed haloes of X-ray emitting gas
through stellar mass loss or accretion from their surroundings is very
difficult to answer. It would be greatly simplified if galaxies with little
or no surrounding intra-cluster medium (ICM) could be observed. The
importance of this issue has recently increased, owing to reports that some
ellipticals may contain very little dark matter or may lack dark matter
haloes entirely \citep{Romanowskyetal03}. This could provide an explanation
for the long known issue of the large degree of scatter in the \LxLbtwo\
relation for elliptical galaxies - galaxies with dark matter haloes have
sufficient mass to prevent the escape of hot gas and/or accrete more, while
those which lack dark matter do not. It is therefore important to observe
ellipticals which do not lie at the heart of a large group or cluster
potential well in order to answer the question of whether ordinary
elliptical galaxies can possess significant quantities of hot gas, without
the aid of aa surrounding deep potential well.

As part of a sample of isolated elliptical galaxies, we have used \chandra\ 
to observe NGC~4555, a fairly luminous elliptical galaxy (log
\LB/\LBsol=10.78) at a distance of $\sim$ 90 Mpc. We find that this
elliptical, which we show is not the dominant galaxy of any group or
cluster, has a sizeable X-ray halo. The hot gas in this halo can be used to
characterise the dark matter halo surrounding the galaxy. Throughout the
paper we assume \Ho=75\kmpspMpc\ and normalise optical B-band luminosities
to the B-band luminosity of the sun, \LBsol=5.2$\times10^{32}$ \ergps.
Abundances are measured relative to the ratios of \citet{GrevesseSauval98},
which differ from the older abundance ratios of \citet{AndersGrevesse79} in
that the solar abundance of Fe is a factor of $\sim$1.4 lower. Details of
the location and scale of NGC~4555 are given in Table~\ref{tab:intro}.

\begin{table}
\begin{center}
\begin{tabular}{lc}
\hline
R.A. (J2000) & 12 35 41.2 \\
Dec. (J2000) & +26 31 23 \\
Redshift & 0.022 \kmps\ \\
Distance (\Ho=75) & 90.33 Mpc \\
1 arcmin = & 26.3 kpc \\
\Dtf\ radius & 19.0 kpc \\
\hline
\end{tabular}
\end{center}
\caption{\label{tab:intro} Location and scale of NGC~4555}
\end{table}

In Section~\ref{sec:obs} we give details of the observation and our data
reduction techniques, and Section~\ref{sec:res} contains the results of our
analysis. Section~\ref{sec:discuss} consists of a discussion of these
results, with particular reference to the issue of dark matter in
early-type galaxies, and Section~\ref{sec:conc} summarizes our results and
conclusions.

\section{Observation and data reduction}
\label{sec:obs}
NGC~4445 was observed with the ACIS instrument during \chandra\ Cycle 3,
Obs ID 2884. A detailed summary of the \chandra\ mission and
instrumentation can be found in \citep{Weisskopfetal02}. The S3 chip was
placed at the focus of the telescope in order to take advantage of the
enhanced sensitivity of the back illuminated CCDs at low energies. The
instrument operated in faint mode, and observed the target for just over 30
ksec. The raw data was reprocessed using \textsc{ciao} v3.0.1 and bad
pixels and events with \asca\ grades 1, 5 and 7 were removed. The data were
corrected to the appropriate gain map, and a correction was made to account
for the time dependence of the gain using the technique described by
Vikhlinin\footnote{http://hea-www.harvard.edu/~alexey/acis/tgain/}. A
background light curve was produced. Some minor background flares were
identified and removed, with all periods where the count rate deviated from
the mean by more than 3$\sigma$ being excluded. The effective exposure of
the observation after cleaning was 23.3 ksec.

Background images and spectra were generated using the blank sky data
described by Markevitch\footnote{http://asc.harvard.edu/cal/}. The data
were cleaned to match the background, and appropriate responses were
created using the \textsc{ciao} tasks \textsc{mkwarf} and
\textsc{mkrmf}. As the ACIS instruments are affected by absorption by
material accumulated on the optical blocking filter, we applied a
correction to the responses. When fitting spectra we generally held the
absorption fixed at the measured galactic value of 1.36$\times10^{20}
cm^{-2}$. Point sources were identified using the \textsc{ciao wavdetect}
tool with a signal threshold of 10${-6}$. We chose to use only the S3 chip
for our analysis, as the galaxy emission should be entirely contained on
S3. This choice of signal threshold means that the detection algorithm
should identify $\leq$1 false source in the field of view. Once identified,
point sources were removed from the data, using regions of twice the radius
given by the detection routine.

\section{Results}
\label{sec:res}
\begin{figure}
\centerline{\epsfig{file=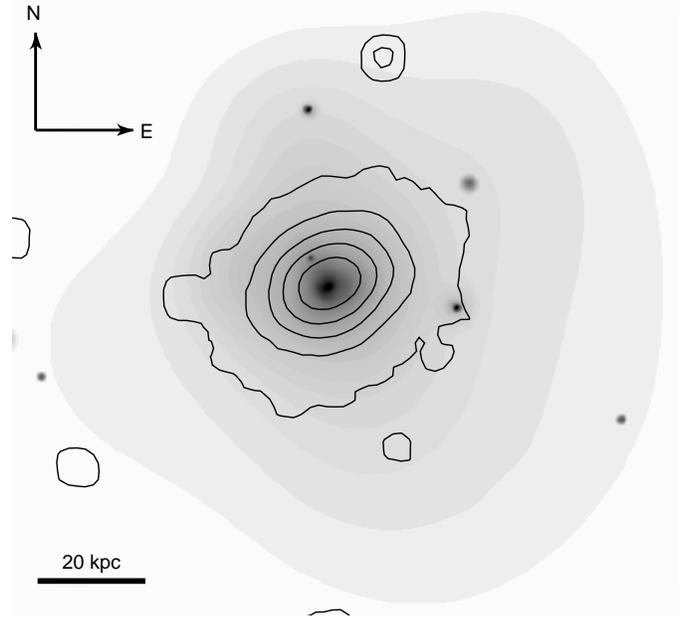,width=9cm}}
\caption{\label{fig:smooth}
Adaptively smoothed image of NGC~4555 created using \textsc{csmooth} with a
signal to noise range of 3 to 5. Optical contours are overlaid to show the
position and extent of the stellar component of the galaxy.}
\end{figure}

We initially prepared adaptively smoothed images of the galaxy, using the
\textsc{ciao} task \textsc{csmooth}. The data were smoothed to show
features with a signal to noise ratio of 3 to 5. A scaled background image
was smoothed and removed from the data and correction for the variations in
exposure across the chip were made. Figure~\ref{fig:smooth} shows the
result of the smoothing with optical contours from the Digitized Sky Survey
overlaid. From this image, it is clear that the X-ray emission extends well
beyond the stellar body of the galaxy, perhaps as far as 60 kpc in some
directions.

\subsection{Two-dimensional surface brightness modeling}
\label{sec:SB}
In order to model the X-ray surface brightness distribution of NGC~4555 we
prepared source and background images in a 0.3-3.0 keV band, with point
sources removed. The energy band was chosen to focus on soft emission and
improve the signal-to-noise of the source. Images were binned to a pixel
size of 1 \arcs. An appropriate exposure map was also generated, and we
used the \textsc{ciao sherpa} package to perform the fitting. As the source
image has many pixels containing few counts (or none), we use the Cash
statistic \citep{Cash79} when fitting. This statistic only provides a
relative measure of the goodness of fit, so that while it allows us to
improve fits and find the best solution for a particular model, it does not
provide an absolute measure of the fit quality. We therefore judge whether
fits are satisfactory (or otherwise) by inspection of azimuthally averaged
radial profiles and residual images. However, the fits are two dimensional,
and so we can determine parameters such as the ellipticity of the halo.

\begin{figure}
\centerline{\epsfig{file=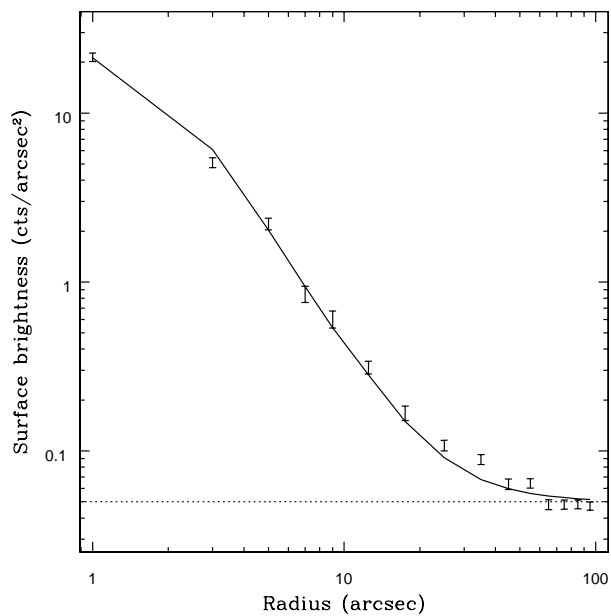, width=9cm}}
\caption{\label{fig:sb}
Azimuthally averaged surface brightness profile of NGC~4555, showing our
best fit beta model. The background level is marked by the dotted line, the
model by a solid line, and the data by error bars. The profile was
azimuthally averaged using elliptical bins with axis ratio and position
angle as in the fit. The radius shown is therefore effectively the minor
axis radius. Note that 1\arcs=0.437 kpc.}
\end{figure}

We performed fits using a beta model, de Vaucouleurs model, and
combinations of the two with a central point source. When combining the de
Vaucouleurs and beta models, the de Vaucouleurs model core radius, axis
ratio and position angle were all fixed at the optically determined values
(r$_c$=16.11\arcs, p.a.=35\deg, axis ratio=1.26), so that the de
Vaucouleurs component would model a discrete source contribution
distributed in the same way as the stellar population. We found that the
beta model provided an adequate fit to the data and inspection of the
residual images and azimuthally averaged radial profiles did not suggest
that more complex models improved the fit. We therefore adopt our best beta
model fit to describe the surface brightness distribution. The fitted
parameters and given in Table~\ref{tab:sb}. Figure~\ref{fig:sb} shows an
azimuthally averaged radial profile with the fitted model.

\begin{table}
\begin{center}
\begin{tabular}{lccc}
r$_{core}$ (\arcs) & \Bfit\ & pos. angle (\deg) & axis ratio \\
\hline\\[-3mm]
1.64$^{+0.15}_{-0.14}$ & 0.577$\pm$0.009 & 26.77$^{+2.43}_{-2.57}$ & 1.12$^{+0.06}_{-0.05}$ \\
\end{tabular}
\end{center}
\caption{\label{tab:sb}
Parameters and 1$\sigma$ errors of out best fitting beta model. Position
angle is measured anti-clockwise from northeast.}
\end{table}

\subsection{Spectral modeling}
\label{sec:spectral}

\begin{figure}
\centerline{\epsfig{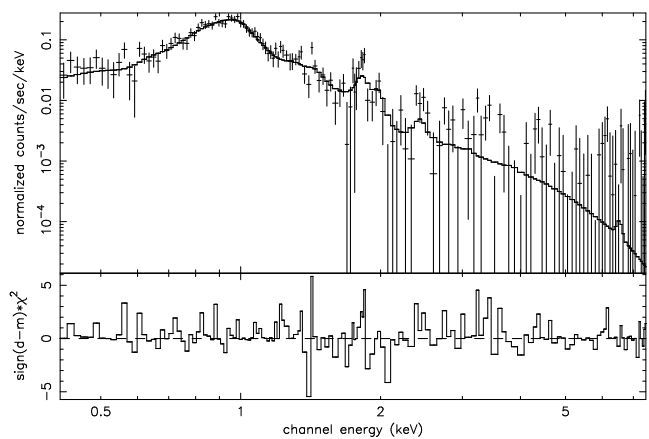}}
\caption{\label{fig:spectrum}Integrated spectrum with best fit APEC+APEC+Bremsstrahlung model. Lower panel shows residuals to the fit plotted in terms of contributions to $\chi^2$.}
\end{figure}

As an initial step we chose to fit the integrated spectrum of the entire
galaxy halo. The spectrum was extracted from a circular region centred on
the peak of the emission, with radius 150\arcs\ ($\sim$65 kpc). After
removal of regions corresponding to point sources were removed, source and
background spectra were extracted, appropriate responses created and
corrected, and the source spectrum grouped to 20 counts per bin. The
spectra were fitted using \textsc{xspec} v11.1.0, ignoring energies lower
than 0.4 and greater than 8.0 keV. The spectrum was fitted fairly
successfully with a model consisting of a 7 keV bremsstrahlung component,
and a hot plasma component, modelled using either the MEKAL
\citep{Liedahletal95,Kaastramewe93} or APEC \citep{Smithetal01} codes. The
bremsstrahlung component was intended to model emission from unresolved
point sources. Both bremsstrahlung and power law models have been
successfully used in this role in previous studies
\citep{IrAtheyBreg03,IrwinSarBreg00}. Results of these fits are given in
Table~\ref{tab:totalfit}.

\begin{table}
\begin{center}
\begin{tabular}{lccc}
 & MEKAL & APEC & APEC+APEC \\
kT & 0.91$\pm$0.04 & 0.95$\pm$0.04 & 0.82$^{+0.04}_{-0.19}$ \\[+1mm]
kT2 & - & - & 1.36$^{+0.55}_{-0.34}$ \\
Z$_{avg}$ & 0.18$^{+0.50}_{-0.18}$ & 0.24$^{+0.49}_{-0.22}$ & 0.52$^{+3.89}_{-0.51}$ \\[+1mm]
Si & 0.52$^{+0.67}_{-0.26}$ & 0.58$^{+0.49}_{-0.31}$ & 1.12$^{+4.91}_{-0.07}$ \\[+1mm]
Fe & 0.41$^{+0.20}_{-0.11}$ & 0.50$^{+0.24}_{-0.19}$ & 1.06$^{+0.69}_{-0.37}$ \\
red. $\chi^2$ & 1.093 & 1.056 & 1.040\\
d.o.f. & 162 & 162 & 160 \\
Flux & 4.86$\times10^{-13}$ & 4.75$\times10^{-13}$ & 4.54$\times10^{-13}$ \\
$f_{bremss}$ & 0.186 & 0.147 & 0.126 \\
$L_{X,gas}$ & 4.75$\times10^{41}$ & 4.64$\times10^{41}$ & 4.43$\times10^{41}$ \\
\end{tabular}
\end{center}
\caption{\label{tab:totalfit} Parameters for our best fits to the
  integrated spectrum. Temperature and abundance are given in terms of 90\%
  error bounds. Flux is calculated for the 0.4-8.0 keV band, is unabsorbed,
  and is given in units of \ergpspcmsq. Gas luminosity is calculated from flux assuming a distance of 90.33 Mpc, and is given in units of \ergps. The fits were performed with \NH\ fixed
  at the galactic value (1.36$\times10^{20}$ cm$^{-2}$) and included a 7 keV
    bremsstrahlung model. $f_{bremss}$ gives the fraction of the total flux
    originating from this component. Z$_{avg}$ gives the abundance of all
    other metals apart from Fe and Si.} 
\end{table}

Although the abundances are rather poorly constrained, freeing Fe and
particularly Si greatly improves the fits. A MEKAL + bremsstrahlung model
with all metals in solar ratios has a reduced $\chi^2$ of 1.196 for 164
degrees of freedom, a significantly poorer fit. Assuming the abundances
given by the APEC fit, we estimate the masses of Iron and Silicon in the
central $\sim$65 kpc of the halo to be M$_{Fe}$=1.49$\times$10$^5$\Msol\ 
and M$_{Si}$=1.83$\times$10$^5$\Msol\ respectively. This gives an
Iron-mass-to-light ratio of only 2.4$\times$10$^{-6}$ \ML, considerably
lower than the values found for galaxy clusters \citep{Finoguenovetal00},
or even the values suggested for small galaxy groups
\citep[4.5$\times$10$^{-4}$,][]{Renzinietal93}. It seems reasonable to
expect that this relatively small quantity of Iron could be produced by the
stellar population of the galaxy without external enrichment.

We also attempted to fit the spectrum with more complex models, including a
two-temperature plasma with bremsstrahlung or power law, cooling flow
models and multi-temperature models. None of these produced a statistically
significant improvement in the fit. However, inspection of the spectrum
showed that the use of a two-temperature plasma model produced a slightly
better fit to the high energy side of the 1 keV Iron peak, and we therefore
record this fit in Table~\ref{tab:totalfit}. Fitting a single temperature
model to spectra from gas with multiple temperature components around 1 keV
is known to produce residuals on both sides of the Iron peak
\citep{Buotefabian98,Buote00b}, so this apparent improvement in fit
probably indicates that NGC~4555 has a multi-temperature halo. It is
notable that the measured abundances for this fit are considerably higher
that those found for the single temperature plasma fits, again agreeing
with previous studies comparing single- and two-temperature fits of
multi-temperature gas. However, the abundances are rather poorly
constrained. To examine whether the large errors in the bins above 3~keV
might be responsible for this lack of precision, we rebinned the spectrum
above 3~keV to have at least 20 counts per bin after background
subtraction. Unfortunately this does not improve the situation; the errors
on temperature and abundance are even larger, and the fit has a reduced
$\chi^2$ of 1.103 (for 85 d.o.f.), not a significant improvement on our
previous fits.

\begin{figure}
\centerline{\epsfig{file=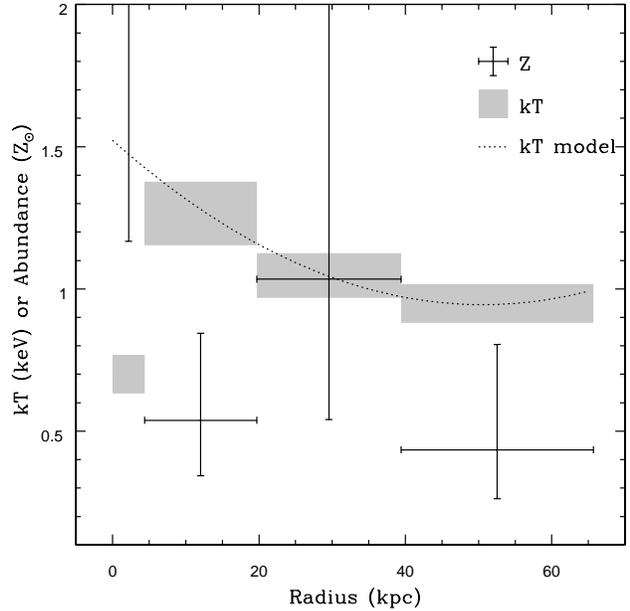, width=9cm}}
\caption{\label{fig:prof}
  Best fit projected temperature and abundance for four radial bins in the
  halo of NGC~4555. Grey boxes mark the temperature of each bin, and
  crosses the abundance, with 90\% error bounds. The dotted line shows a
  fitted model of the temperature profile, ignoring the central bin, as
  described in Section~\ref{sec:mass}.}
\end{figure}

We have sufficient detected counts from the galaxy to allow us to split the
halo into four bins and fit spectra from these individually. We use
elliptical annuli with axis ratio and position angle taken from the best
fitting surface brightness model. The angular sizes of the bins are
0-20\arcs, 20-45\arcs, 45-90\arcs\ and 90-150\arcs, with distances measured
on the semi-minor axis. The spectra and associated responses and background
spectra were prepared as described above and fitted individually. These
spectra were not of the quality required to fit individual metal lines, so
we used MEKAL models with bremsstrahlung components added if necessary.
Figure~\ref{fig:prof} shows the fitted temperature and abundance in each
bin. In each bin we hold the hydrogen column fixed at the galactic value.
Only the central bin was improved by the addition of a bremsstrahlung
component, which contributes $\sim$16 per cent of the emission in this bin.
A bremsstrahlung contribution is not ruled out in the other bins, but
provides no significant improvement in the fit. We also note that inclusion
of such a component does not significantly affect the fitted temperature of
the MEKAL component. Fit quality in the central bin is good, reduced
$\chi^2$=0.87 for 29 degrees of freedom. Fit quality in the outer bins is
considerably poorer, with reduced $\chi^2 \sim$1.3 in bins 2 and 3 and 1.13
in the outermost bin. We tried numerous models including MEKAL and APEC
plasmas, power law and bremsstrahlung components, and multi-temperature
components such as CEMEKL and MKCFLOW. None produced any significant
improvement over a single temperature plasma, and we therefore use the
parameters determined by these models as the best fit. We also tested our
background subtraction, using a local background extracted from a
source-free region of the S3 chip. Bins 1 and 2 showed no significant
change in fit statistic or model parameters when using the local
background. The best fit models for bins 3 and 4 had larger errors on
temperature and abundance but were consistent with the fits performed using
the blank-sky background data. Fit statistics for these bins were also slightly
altered, presumably owing to the poorer statistics of the local background
data, which is the likely cause of the larger error regions. We conclude
from this that our use of the blank-sky background data is justified and
gives accurate results.

Figure~\ref{fig:prof} shows that temperature rises from the outer parts
toward the core, but turns over and falls in the core. This suggests that
radiative cooling in the dense core of the galaxy halo is effective, and
has significantly reduced the mean temperature of the gas. The abundance in
each bin is more poorly defined, but appears to show a general decrease
with increasing radius. This is consistent with enrichment of the galaxy
halo by metals lost from the stellar population through stellar winds and
supernovae.

\subsection{Mass, Entropy and cooling time}
\label{sec:mass}
Given the surface brightness and temperature modelling of the halo of
NGC~4555, it is possible to estimate three dimensional properties such as
mass, entropy and cooling time. The density profile of the gas can be
estimated from the measured profiles and then normalised to reproduce the
X-ray luminosity of the galaxy, determined from our best fitting APEC
model. Given the density profile we can use the well known equation for
hydrostatic equilibrium,

\begin{equation}
M_{tot}(<r) = -\frac{kTr}{\mu m_pG}\left(\frac{d{\rm ln}\rho_{gas}}{d{\rm
      ln}r}+\frac{d{\rm ln}T}{d{\rm ln}r}\right),
\end{equation}

\noindent to calculate the total mass within a given radius. From as
density and total mass, we can calculate parameters such as gas fraction,
cooling time and entropy where entropy is defined to be

\begin{equation}
S = \frac{T}{n_e^{\frac{2}{3}}}.
\end{equation}

We estimate the errors on the derived values using a monte-carlo
technique. The known errors on the temperature and surface brightness
models, and on other factors such as the total luminosity, are used to
randomly vary the input parameters. We then generate 10000 realisations of
the derived parameters profiles, and use these to calculate the 1$\sigma$
error on each parameter at any given radius.

\begin{figure*}
\centerline{\epsfig{file=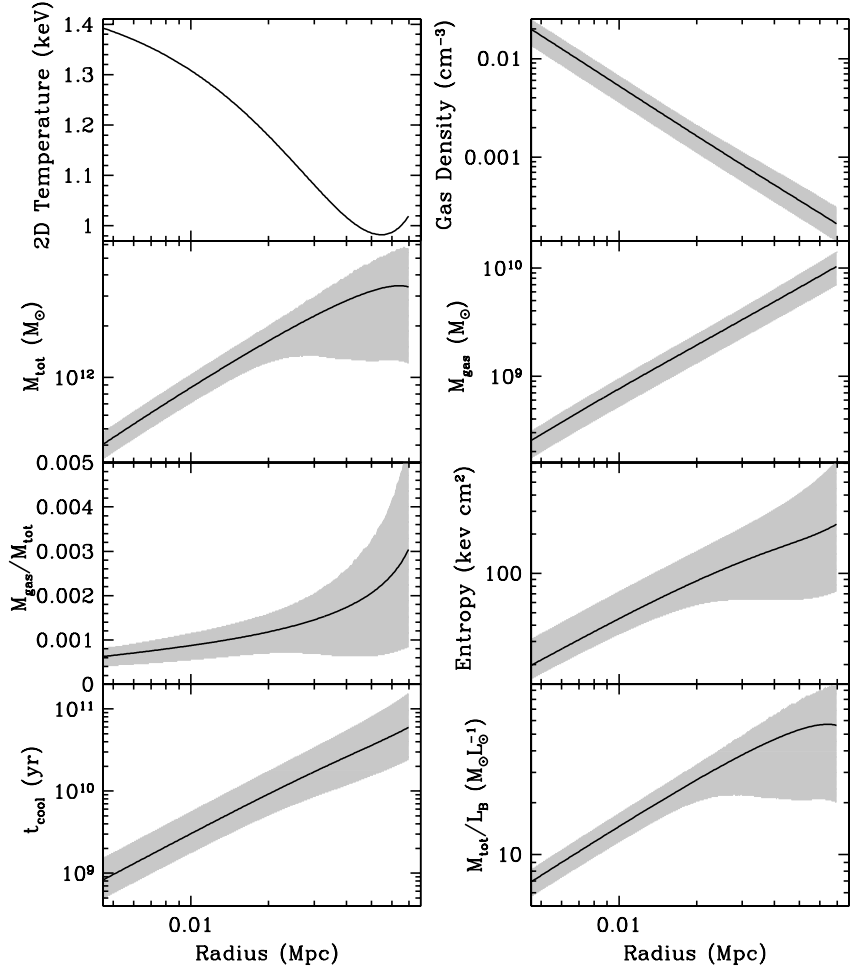,width=12cm}}
\caption{\label{fig:3d}
  Projected temperature and deprojected gas density, gravitational mass,
  gas mass, gas fraction, entropy, cooling time and mass-to-light ratio for
  NGC~4555. The inner boundary of the plots is 5 kpc, the radius within
  which we know our temperature model to be inaccurate. Solid lines show
  values derived from the best fitting temperature and surface brightness
  models, grey regions show 1$\sigma$ errors.}
\end{figure*}

The only issue which arises in these calculations is the question of how
well we can model the temperature profile, which declines sharply in the
galaxy core. The central bin has no effect on the value of the parameters
at larger radii and increases the complexity of the required model, and we
therefore choose to exclude the central temperature bin, and model the
temperature profile as if there were no central cooling. The remaining
three bins can be well described with a quadratic, which is shown as a
dotted line in Figure~\ref{fig:prof}. Based on this model of the
temperature we calculate the parameters shown in Figure~\ref{fig:3d}.

In order to calculate the mass-to-light ratio, we use the H band near
infra-red surface brightness profile of \citet{Gavazzietal00}. This profile
is made up of a bulge component described by a de Vaucouleurs profile, with
$r_e$=4.01\arcs, and an exponential disk with $r_e$=17.42\arcs. We assume
that these measurements hold for the B band optical light, and normalise
the profile to the B band luminosity, with a bulge-to-disk ratio of 0.33.
We note that the mass-to-light ratio at the inner limit of the plot is
$\sim$9, a little higher than the value of 5-8 generally assumed for
the mass-to-light ratio of the stars alone.

The outer limit of the plots is determined by the radius to which we can
measure the temperature, and the inner limit by the radius of the innermost
temperature bin. The inner limit is chosen to exclude the 5 kpc radius
region in which we know our temperature model to be wrong. We also note
that both the temperature and surface brightness model profiles were
determined using the semi-minor axis as the radius descriptor. The profiles
shown in Figure~\ref{fig:3d} should therefore be considered to show
properties which are azimuthally averaged around ellipses whose position
and ellipticity are determined by the best fitting surface brightness
model, with the radial scale indicating the semi-minor axis.

\subsection{Point sources}
\label{sec:ps}
As described in Section~\ref{sec:obs}, we used the \textsc{ciao wavdetect}
tool to identify point sources in the field of view. Several sources are
found to lie within the extended emission surrounding NGC~4555, but only
two lie within optical extent of the galaxy, defined by the \Dtf\
ellipse. Of these, one is coincident with the galaxy core. Surface
brightness fitting of the galaxy X-ray halo (see Section~\ref{sec:SB}) does
not show strong evidence for a central point source, and we conclude that
the identification from \textsc{wavdetect} actually corresponds only to the
peak of the galaxy halo, not to a separate source.

The remaining source lies $\sim$15\arcs\ NE of the galaxy centre, and does
not correspond to any feature visible in the DSS optical or 2MASS infra-red
images. There are no objects listed at its position in NED. We extracted a
count rate for the source, in the full \chandra\ band and in a 0.5-2.0 keV
band. Background subtraction was carried out using a region immediately
surrounding the source, and should therefore account for contamination by
the galaxy halo. Based on the background subtracted count rate we estimated
the flux and luminosity of the source assuming a bremsstrahlung model with
kT=5 keV, and a power law model with $\Gamma$=1.96, typical of high
luminosity point sources in other galaxies \citep{IrAtheyBreg03}. The
results are given in Table~\ref{tab:ps}.

\begin{table}
\begin{center}
\begin{tabular}{lc}
\hline
R.A. & 12:35:41.74 \\
R.A. error (\arcs) & $\pm$0.127 \\
Dec. & +26:31:35.2 \\
Dec. error (\arcs) & $\pm$0.123 \\
Counts (0.1-12.0 keV) & 15.04 \\
Counts (0.5-2.0 keV) & 15.76 \\
Flux (BR, \ergpspcmsq) & 2.719$\times$10$^{-15}$ \\
Flux (PL, \ergpspcmsq) & 2.768$\times$10$^{-15}$ \\
\hline
\end{tabular}
\end{center}
\caption{\label{tab:ps}
Properties of the off centre point source in NGC~4555. Fluxes are
calculated assuming no absorption, a 0.5-2.0 keV energy band and either a 5
keV bremsstrahlung (BR) model or a power law (PL) model with $\Gamma$=1.96.}
\end{table}

If this source is in fact part of the NGC~4555 system and is actually a
single object rather than an unresolved cluster of sources, then it is
extremely luminous, log \Lx$\sim$39.43 \ergps. We can estimate the
probability of finding a background source with the measured flux within
the \Dtf\ ellipse based on the sources found in the \chandra\ deep field
south \citep{Tozzietal01}. The \Dtf\ ellipse has an area of
$\sim$3.62$\times$10$^{-4}$ square degrees, which means that we would
expect to find $\sim$0.02 sources with the observed (or greater) flux in
that area. Assuming that the background distribution of sources is the same
as the \chandra\ deep field, this suggests that the source is probably part
of NGC~4555, and must therefore be classed either as an ultra-luminous X-ray
source (ULX), or considered to be an unresolved cluster of sources.

\section{Discussion}
\label{sec:discuss}

\subsection{Environment}
\label{sec:env}
NGC~4555 was selected as part of a sample of isolated ellipticals,
extracted from the Lyon-Meudon Extragalactic Data
Archive\footnote{http://leda.univ-lyon1.fr} (LEDA). Sample galaxies were
selected using the following criteria:
\begin{enumerate}
\item Morphological type $T\leq-3$, \ie early-type galaxies.
\item Virgocentric flow corrected velocity $v\leq9000$\kmps.
\item Apparent B-band magnitude $B_T\leq14.0$.
\item Galaxy not listed as a member of a Lyon Galaxy Group
  \citep[LGG,][]{Garcia93}.
\end{enumerate}

The restrictions on apparent magnitude and recession velocity were imposed
to minimise the effect of incompleteness in the catalogue. The LEDA
catalogue is known to be 90 per cent complete at $B_T=14.5$
\citep{Amendolaetal97}, so our sample should be close to 100 per cent
statistically complete. The selection process produced 330 galaxies which
could be considered as potential candidates. These were compared to the
rest of the catalogue and accepted as being isolated if they had no
neighbours which were:

\begin{enumerate}
\item within 700\kmps in recession velocity,
\item within 0.67 Mpc in the plane of the sky,
\item less than 2 magnitudes fainter in $B_T$.
\end{enumerate}

The criteria were imposed to ensure that the galaxies did not lie in groups
or clusters, and to ensure that any neighbouring galaxies were too small to
have had any significant effect on their evolution or properties.

To check the results of this process, all candidate galaxies were compared
to the NASA Extragalactic Database (NED) and the Digitised Sky Survey
(DSS). A NED search of the area within 0.67 Mpc of each candidate
identifies galaxies not listed in LEDA. We also examine DSS images of this
region for galaxies of similar brightness to the target which are not
listed in either catalogue. The process produced 40 candidate isolated
elliptical galaxies, of which NGC~4555 is one.

\begin{figure}
\centerline{\epsfig{file=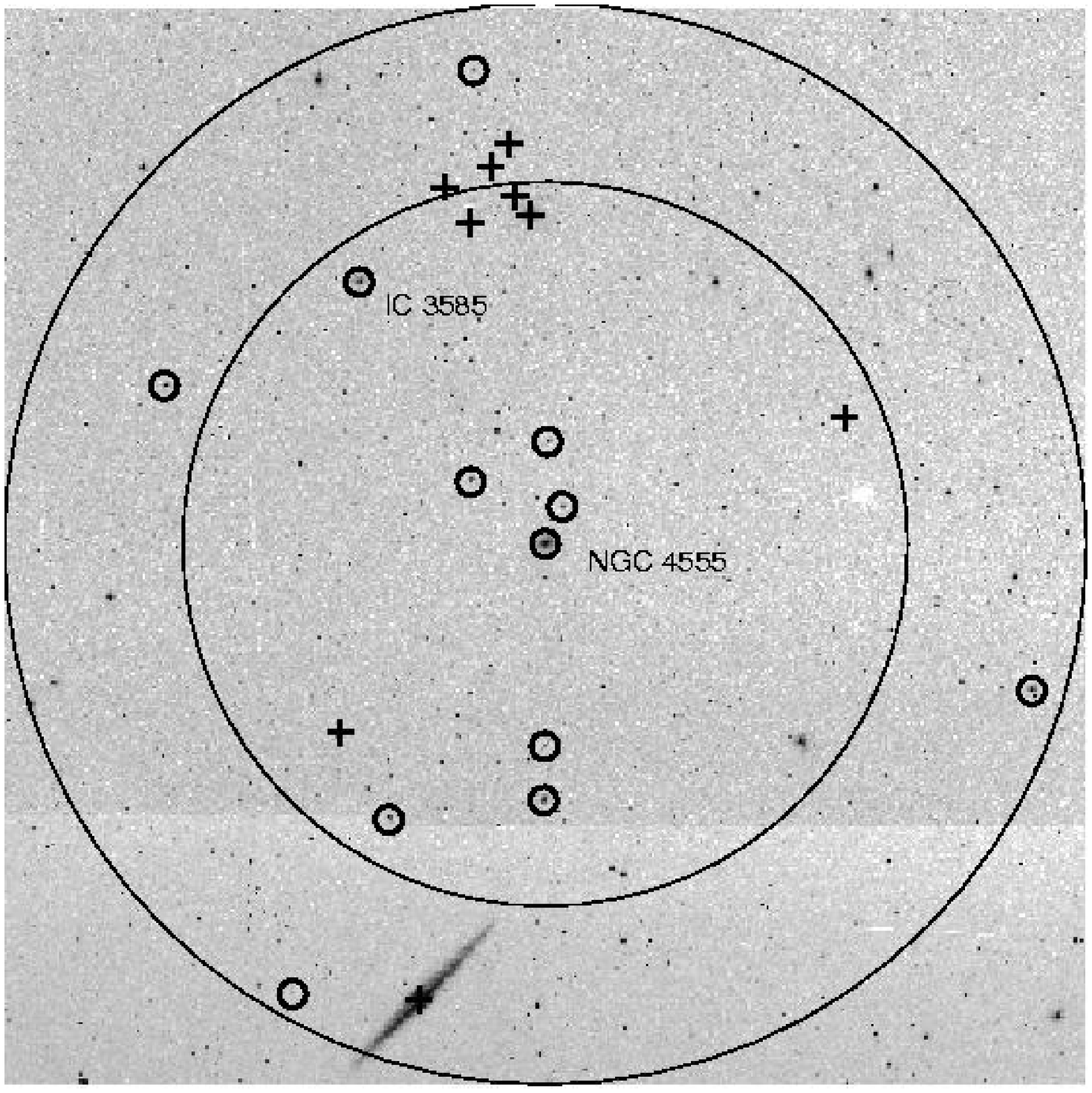,width=8.5cm}}
\caption{\label{fig:env}
  A Digitized Sky Survey image of the environment of NGC~4555 with
  positions of galaxies overlaid. The two large circles mark regions of
  radius 0.67 Mpc and 1 Mpc, centred on NGC~4555. Small circles mark
  galaxies whose recession velocities differ from that of NGC~4555 by
  $<$700 \kmps, crosses those $>$700 \kmps\ away. For further details, see
  the text. North is to the top in this image and west is to the right. The
  linear feature running across the lower third of the figure is the
  junction between DSS plates.}
\end{figure}

Although NGC~4555 meets the criteria described above, it does have a number
of galaxies relatively close to it. We have therefore examined the
surrounding galaxy population in order to determine whether NGC~4555 is on
the outskirts of a group or cluster. Figure~\ref{fig:env} shows a DSS image
of the region surrounding NGC~4555, which is marked in the centre of the
image. Scale is shown by the two large circles, which have radii of 0.67
and 1 Mpc. All galaxies listed in NED within 1 Mpc which have measured
recession velocities are marked on the plot. Crosses indicate galaxies
$>$700\kmps\ away from NGC~4555, circles those within this velocity range.
All galaxies marked by circles have apparent magnitudes at least 2
magnitudes fainter than NGC~4555, with the exception of IC~3585, a nearby
S0 galaxy which NED shows to be $\sim$600 kpc and 687 \kmps\ from NGC~4555.
It is only $\sim$1.5 magnitudes fainter than our target, and so would
appear to violate our isolation criteria. However, the velocities
available from NED have not necessarily been corrected in the same way, so
the LEDA velocities should provide the more accurate measure of relative
distance. Using the LEDA Virgocentric flow corrected velocities, the
difference between the two galaxies is 758 \kmps, putting IC~3585 just
outside our chosen velocity limit.

\begin{figure}
\centerline{\epsfig{file=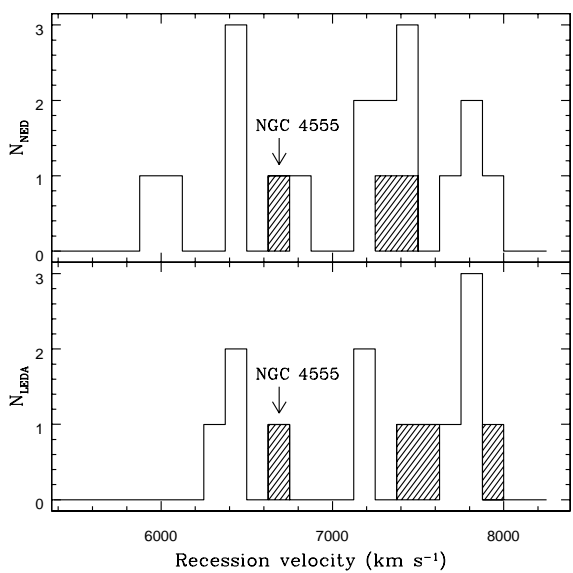,width=8.5cm}}
\caption{\label{fig:histo}
Redshift distribution of the galaxies within 1 Mpc (in the plane of the
sky) of NGC~4555. The upper panel shows galaxies with redshifts listed in
NED, while the lower panel shows galaxies listed in LEDA with both
redshifts and B-band magnitudes. Shaded areas show galaxies whose B-band
luminosity is $>$10$^{10}$ \ergps.}
\end{figure}

Figure~\ref{fig:histo} shows histograms of the local velocity field within
1 Mpc of NGC~4555. In the upper panel we show galaxies found in NED, which
were required only to have a measured redshift. In the lower panel we show
galaxies found in LEDA, which were required to have a measured redshift and
apparent magnitude. We have marked galaxies whose magnitudes indicate that
they have a B-band luminosity in excess of 10$^{10}$ \ergps. The remaining
galaxies have $2\times10^9 < L_B < 10^{10}$ \ergps. NGC~4555 is represented
in each plot by the shaded region at lowest recession velocity.

Although the plots do suggest some sort of extended structure, it is clear
that the galaxies in this region do not form a relaxed group. Similarly, it
is clear that NGC~4555 is separated from the other galaxies of significant
mass (for which \LB\ is a proxy) by significant distance and velocity
difference. It seems possible that NGC~4555 is part of a small filament of
galaxies, or that it (with some small companions) is falling into a very
poor group to the north. It is very unlikely that NGC~4555 is the dominant
galaxy of a virialised group, and it therefore must be assumed that it does
not lie in the centre of a massive group-scale dark matter halo.

NGC~4555 does not appear in the sample of ``very isolated early-type
galaxies'' defined by \citet{Stockeetal04}. This sample is drawn from the
\citet{Karachentseva73} isolated galaxy catalogue, which was compiled by
examining images of the environment of all bright ($m_B \leq$ 15.7)
galaxies in the \citet{Zwickyetal68} catalogue. Galaxies were classified as
isolated if they had no companion within 20 galaxy diameters, where a
companion was defined as any galaxy with angular diameter within a factor
of 4 of that of the candidate. Stocke et al have further cleaned the
resulting sample by examining the Palomar Observatory Sky Survey (POSS)
plates to confirm isolation, and performing deeper imaging around a number
of the galaxies. This should lead to a sample with few galaxies falsely
identified as isolated, but will rule out NGC~4555 because of similarly
sized galaxies nearby which are in fact distinct from it in redshift
space. It is possible that NGC~4555 is ruled out by the presence of
NGC~4565, the large edge on spiral galaxy at the lower edge of
Figure~\ref{fig:env}, which has a recession velocity $>$5000 \kmps\ smaller
than NGC~4555.

We also note that NGC~4555 has been identified as a member of a galaxy
group by two previous studies. \citet{Mahtessian98} suggest that it is part
of a triplet with IC~3582 (which is fainter than our magnitude cut off) and
NGC~4556 (which has a recession velocity $\sim$800 \kmps\ than that of
NGC~4555). \citet{Ramellaetal02} use a friends of friends algorithm to
search for associations in the Updated Zwicky Catalog \citep{Falcoetal99}
and Southern Sky Redshift Survey \citep{DaCostaetal98} galaxy catalogues.
They suggest that NGC~4555 is part of a group of 7 objects with a range of
recession velocities (6615-7819 \kmps). They calculate the group velocity
dispersion to be 568 \kmps\ and from this estimate the virial radius (1.64
Mpc), total mass (3.7$\times$10$^{14}$ \Msol) and mass-to-light ratio
($\sim$1740 \ML). These values seem rather large for a system so poor in
galaxies, and are particularly surprising given the lack of a group X-ray
halo. From the mass-temperature relation of clusters and groups
\citep{SandersonPonman03} we would estimate that a system of this mass
would have an X-ray temperature of kT$\sim$8 keV, and it seems unlikely
that such a large dark matter potential could have failed to accrete some
gas during its formation. We therefore consider it likely that the
identification of NGC~4555 and its neighbours as a collapsed group is
faulty, and that they at best form a loose association, not dominated by a
single large dark matter halo.

\subsection{Comparison with other systems}
As mentioned in Section~\ref{sec:intro}, recent studies of the velocity
dispersion profiles of some ``ordinary'' ellipticals have shown them to
have lower mass-to-light ratios than were expected
\citep{Romanowskyetal03}. This suggests that the dark matter content of
these galaxies is minimal, at least within the radius to which the
measurements extend ($\sim$5$r_e$). The profiles are incompatible with the
dark matter profiles predicted by simulations of structure formation in a
cold dark matter dominated universe, leading to the conclusion that either
the haloes take the predicted form but have much lower masses than expected
(the results are consistent with zero dark mass), or that the dark matter
halo takes a very different form, perhaps with most of the mass at larger
radii. Four elliptical galaxies, NGC~821, NGC~3379, NGC~4494 and NGC~4697,
have been shown to have this unexpected lack of dark matter. Of these,
three are in small groups and are the brightest elliptical in each
system. NGC~821 is a rather more isolated system, not part of any known
group or cluster. All of the galaxies are relatively X-ray faint,
undetected (or only marginally detected, in the case of NGC~4697) by
\rosat, and two have been shown to possess relatively small amounts of
X-ray emitting gas \citep{IrwinSarBreg00,OSullivanPonman04}. This lack of a
sizeable X-ray halo could be a consequence of the lack of dark matter in
these galaxies, as the stellar mass alone would be insufficient to retain a
large gaseous halo.

For comparison with NGC~4555, we use the mass-to-light ratios at 5$r_e$ for
NGC~4494, NGC~3379 and NGC~821, quoted by \citet{Romanowskyetal03}. An
overall mass-to-light ratio for NGC~4697 of M/L=11\ML\ was found by
\citet{Mendezetal01}, but this is for a radius of $\sim$3$r_e$. For
NGC~4555 itself we have three different measurements of $r_e$, one overall
value, which assumes a de Vaucouleurs profile, and the effective radii of
bulge and disk components from \citet{Gavazzietal00}, who decompose the
surface brightness profile of the galaxy into two components. Their result
is rather confusing, as their profiles assign only 33 per cent of the H
band light of the galaxy to the bulge component, which would suggest that
NGC~4555 is a misclassified S0. However, NGC~4555 is classified as an
elliptical in both NED and LEDA, and in fact Gavazzi et al. list its
morphological type as elliptical. We therefore believe that the overall
effective radius probably gives the best indication of the scale of the
galaxy, but quote mass-to-light ratios at three radii. The ratios are
listed in Table~\ref{tab:ML}. Using the overall effective radius, it is
clear that NGC~4555 has a considerably larger mass-to-light ratio than the
three other ellipticals. The Gavazzi et al disk component effective radius
produces a similar result, and only if the bulge component is used do we
find comparable figures. We assume that this is a poor measure of the true
scale of the NGC~4555, and that the galaxy has a larger mass-to-light ratio
than the three Romanowsky et al. ellipticals.late

\begin{table}
\begin{center}
\begin{tabular}{llc}
Galaxy & Radius & M/L \\
 & (\arcs) & (\ML) \\ 
\hline
NGC~821 & 5$r_e$=250 & 13-17 \\
NGC~3379 & 5$r_e$=175 & 5-8 \\
NGC~4494 & 5$r_e$=245 & 5-7 \\
\hline
NGC~4555 & 5$r_e$=80.6 & 43.6 \\
         & 5$r_{eb}$=20.1 & 13.8 \\
         & 5$r_{ed}$=87.1 & 46.4 \\
\end{tabular}
\end{center}
\caption{\label{tab:ML}
Mass-to-light ratios for NGC~4555 and three ellipticals believed to have
little or no dark matter
. $r_e$ is the effective
radius of the galaxy as a whole, $r_{eb}$ and $r_{ed}$ are the effective
radii of the bulge and disk components of NGC~4555.}
\end{table}

Another class of elliptical galaxies which should be compared with NGC~4555
are the fossil groups. These are systems in which all the major galaxies of
a group collapse at an early epoch, merging to form a single giant
elliptical embedded in a group scale dark matter and X-ray halo
\citep{Ponmanbertram93}. Because these systems appear as a single
elliptical with no neighbouring galaxies of similar size, our optical
selection criteria would likely identify them as candidate isolated
ellipticals. Fossil groups can be identified by four main features
\citep{Jonesetal03}. These are: 1) They have an X-ray luminosity typical
for a galaxy group (\Lx\ $> 10^{42}$ \ergps), 2) Their X-ray halo is highly
extended, 3) The dominant elliptical is at least 2 magnitudes brighter than
the other group galaxies, and 4) The dominant elliptical is surrounded by a
halo of faint galaxies, the unmerged members of the galaxy group. Criteria
3 and 4 can be refined in that a luminosity function of the galaxies in the
fossil group will have only one galaxy with optical luminosity greater than
L$^*$, the luminosity of this galaxy will be significantly higher than
would be expected from the luminosity functions of other groups, but the
tail of dwarf galaxies will be relatively similar to that of other galaxy
groups \citep[see in particular Fig. 7,][]{Jonesetal00}. NGC~4555 fails the
first of these criteria in that its X-ray luminosity is only
$\sim$4.4-4.7$\times10^{41}$ \ergps\ (model dependent). The extension of
the halo of NGC~4555 is also smaller than any known fossil group. The one
possible exception is NGC~6482, a fossil group observed with \chandra,
whose halo extends off the ACIS-S detector and so is as yet poorly
characterised as regards extent \citep{Khosroshahietal04}. The optical
luminosity and isolation of NGC~4555 has been dealt with in
Section~\ref{sec:env}, and from this the galaxy appears to be isolated
enough to meet condition 3 for fossil group status. Testing condition 4 is
more difficult, as we do not have redshifts for the faint galaxies around
NGC~4555. However, we extracted the positions of all galaxies without
redshift listed in NED within 0.444\deg\ of NGC~4555 (equivalent to 0.7 Mpc
at the distance assumed, 90.33 Mpc) and plotted a radial profile of surface
number density of galaxies around our isolated elliptical. If NGC~4555 were
a fossil group we might expect to see higher number densities around it,
caused by the surrounding halo of fainter group members. We note that the
galaxies listed have magnitudes as faint as 19.5 mag. and we are therefore
likely to be missing some of the faintest galaxies which have not been
identified and listed in NED. However, we find no evidence of an
overdensity of faint galaxies around NGC~4555 and it seems unlikely that
the inclusion of a small number of fainter objects could change this
result. This combination of environmental and X-ray properties argues
strongly against NGC~4555 being a fossil group.

It is also possible to compare our results with other X-ray mass estimates
from the literature. A number of early-type galaxies have mass estimates
available, but the requirement for high quality data means that the
majority of these galaxies are highly luminous and reside in the cores of
groups and clusters. Techniques similar to ours have been used to produce
mass profiles of galaxies such as NGC 507 \citep{Paolilloetal03}, NGC~1399
and NGC~1404 \citep{Paolilloetal02}, NGC~2563, NGC~4325 and NGC~2300
\citep{Mushotzkyetal03}, NGC~4472, NGC~4636 and NGC~5044 \citep[][and
references therein]{MathewsBrighenti03}.  However, these are all group or
cluster dominant galaxies, with the exception of NGC~4472, which dominates
a subclump in the Virgo cluster, and NGC~1404, which has a truncated halo,
probably caused by interaction with Fornax cluster gas. Using an alternate
technique, \citet{LoewensteinWhite99} determined constraints on mass and
mass-to-light ratio for a sample of $\sim$30 galaxies for which global
temperature measurements are available, but more detailed density and
temperature profiles are not. The authors find that the relation between
X-ray temperature and optical velocity dispersion determined from the
sample of \citet{DavisWhite96} implies a fairly constant mass-to-light
ratio within 6$r_e$ of $\sim$23 $h_{75}$ \ML. Once again, however, the
sample of galaxies used is heavily weighted toward those ellipticals which
dominate groups and clusters - 20 of the 30 galaxies listed by Davis \&
White are group or cluster dominant ellipticals, and a further 3 should
probably be excluded -- M32, a dwarf elliptical interacting with M31,
NGC~4406, undergoing strong interaction with the Virgo ICM, and NGC~4472
which is mentioned above. It is also worth noting that of the remaining 7
galaxies, 6 are located in either the Virgo or Fornax clusters, and could
have had their dark matter halos altered by interactions with their
environment.

A third technique for determining the presence of dark matter from X-ray
observations has been demonstrated in a series of papers studying NGC~1332,
NGC~3923 and NGC~720
\citep{Buoteetal02,BuoteCanizares98,BuoteCanizares97,BuoteCanizares96,BuoteCanizares94}.
These galaxies have non-spherical stellar and X-ray distributions, and the
authors employ geometric arguments to show the need for a dark matter
component to explain their ellipticity and the relative position angles of
the gas and stars. They are able to model the dark halo and produce strong
constraints on the total mass and mass-to-light ratio, under the
assumptions of hydrostatic equilibrium and that rotation has a minimal
influence on the potential. These models suggest that NGC~720 has a
mass-to-light ratio of $\sim$19 \ML\ at 3$r_e$, NGC~3923 has M/L=17-32 \ML\ 
at $\sim$47 $h_{75}$ kpc, and NGC~1332 has M/L$\sim$31-143 \ML\ within
$\sim$46 $h_{75}$ kpc. However, these galaxies are all group dominant
ellipticals, and while the groups are not as X-ray luminous as objects such
as NGC~5044, they are gravitationally bound systems, as confirmed by the
group-scale X-ray haloes around NGC~720 and NGC~3923
\citep{Mulchaeyetal03}. The mass-to-light ratios for these galaxies are in
general agreement with those found by the other methods described above,
suggesting that at the radii of interest to us ($\sim$5-6$r_e$), most
ellipticals for which X-ray estimates are available have mass-to-light
ratios of at least 20 \ML, and in some cases considerably higher. At larger
radii, the mass-to-light ratio can rise considerably higher, above 100\ML,
indicating the influence of the group or cluster dark matter halo.

The results of these X-ray mass studies show that dark matter is present in
and around the galaxies observed. This is an important result, but the fact
that these galaxies are almost all in the cores of much larger systems must
raise the issue of whether the dark matter profiles derived for them are
really describing the galaxy, the group halo, or some combination of the
two. The galaxies for which individual mass profiles or determinations are
available can tell us a great deal about ellipticals at the centres of larger
structures, but we cannot know what the influence of their environment
is. There are a small number of objects in the \citet{LoewensteinWhite99}
study which might be individually useful, but unfortunately the technique
used applies to the full sample of galaxies and does not provide
mass-to-light ratios for each elliptical. We therefore conclude that
while we can compare our results for NGC~4555 to these X-ray mass
estimates, we must do so cautiously, considering that we may be comparing
systems of quite different scale and content.


Two conclusions might be drawn from these comparisons. Firstly, as we have
already demonstrated, NGC~4555 is not in the core of a virialised group,
and is therefore not surrounded by a group or cluster scale dark matter
potential. Its observed properties therefore confirm that elliptical
galaxies can possess dark matter haloes of their own, regardless of their
environment. The mass of dark matter seems to be comparable to that found
for ellipticals in he cores of groups and clusters. It is to be expected
that at larger radii, group and cluster dominant ellipticals would have
considerably higher mass-to-light ratios that NGC~4555, and it would be
interesting to extend our mass profile further to investigate this. The low
dark matter content found in the three ellipticals studied by Romanowsky
\etal\ show that while ellipticals can possess dark haloes, not all of them
do. This raises an important question; what determines whether elliptical
galaxies have dark matter haloes?

One possibility is that all elliptical galaxies are formed with dark matter
haloes, but some later lose them through interactions with other
galaxies. Close interactions between galaxies can cause tidal stripping of the
dark matter halo \citep{Mathewsbrighenti97} as well as gas and
stars. Simulations of interactions among multiple galaxies in a compact
group suggest that a large amount of the dark matter may in fact be
dispersed, forming a common halo but not bound to any particular galaxy
\citep{Barnes89}. The simulations also show that the dominant galaxy of the
group retains a sizeable dark matter halo, but it is possible that there
are circumstances in which this would not be the case. NGC~4494, NGC~3379
and NGC~4697, all of which are the most luminous and presumably most
massive elliptical in their groups, might have lost their dark matter in
this way.

NGC~821 is more difficult to explain as a product of tidal stripping. The
galaxy is relatively isolated, not a member of any group or cluster, and in
fact meets the isolation criteria described in Section~\ref{sec:env}. As it
has no massive neighbours and does not appear to be part of a larger
structure, there is little chance that it has suffered tidal
stripping. Another method of removing its dark matter halo, or forming such
a galaxy without a halo, therefore appears to be required.

Whereas we can compare the mass and mass-to-light ratio of NGC~4555 with
the galaxies studied by Romanowsky \etal, comparison of the gas mass and
gas fraction with other systems is hampered by the lack of detailed studies
of bright ellipticals outside the cores of groups and clusters. As
mentioned previously, almost all elliptical galaxies for which mass and gas
mass profiles have been calculated are the dominant galaxies of larger
structures. Their properties are likely to be affected by the surrounding
potential, and models of galaxies embedded in a dense intra-group medium
show that the galaxy can accrete gas from its environment
\citep{BrigMath99}, thereby changing its gas fraction. An alternative is to
compare NGC~4555 to a sample of poor groups. These have gas temperatures
similar to that we observe in NGC~4555, and sufficient numbers have been
studied to make samples reliable. We use the 0.3-1.3 keV sample of
\citet{Sandersonetal03} which contains two elliptical galaxies but is
dominated by poor groups. 

For an accurate comparison, it is necessary to compare properties at a
common radius, relative to the overall size of the system. This is usually
done using \Rth\ (where \Rth\ is a good approximation
of the virial radius). Unfortunately we do not have a measured value of
\Rth\ for NGC~4555 and we cannot calculate one based on our three
dimensional models, as the temperature model is not accurate beyond the
outer radius we have used. Based on the Sanderson \etal groups, a typical
\Rth\ for a system of this temperature might be $\sim$500 kpc. If we adopt
this value, we see that while the 0.3-1.3 keV poor groups have gas
fractions of $\sim$1 per cent at 0.1$\times$\Rth, NGC~4555 has a gas
fraction a factor of $\sim$5 lower. This result is of course dependent on
the value of \Rth\ chosen. The Sanderson \etal systems have values of \Rth\
ranging from $\sim$200-800 kpc, but even if we assume a larger \Rth\ ( and
therefore measure gas fraction at a larger radius) we still find that
NGC~4555 has a gas fraction considerably lower than that of poor groups.

\section{Summary and conclusions}
\label{sec:conc}
We have used \chandra\ to observe the relatively isolated early-type galaxy
NGC~4555. An examination of its environment suggests that the galaxy is not
a member of a virialised group, though it may be part of a loose
association or filament of galaxies. It is unlikely that the galaxy is in
the core of a larger, group-scale dark matter potential, as is the case
with many of the early-type galaxies whose X-ray properties have been
studied in more detail. Despite the lack of a surrounding group potential,
we find that the galaxy possesses an extended gaseous halo with a
temperature of kT$\sim$0.95 keV and Fe abundance $\sim$0.5\Zsol. 

We measure the surface brightness distribution of the gaseous halo and find
that it is reasonably well described by a single beta model, supporting the
spectral results which suggest that emission from gas dominates over
emission from point sources. We also measure the temperature and abundance
profiles and find evidence for a central cooling region, though
confirmation of this would require fitting a multi-temperature model which
at present we have insufficient counts to do. Assuming the gaseous halo to
be in hydrostatic equilibrium, we use these measurements to estimate the
gas mass, entropy, cooling time, total gravitating mass and mass-to-light
ratio of the system. We find a mass-to-light ratio of
42.7$^{+14.6}_{-21.2}$ \ML\ at 5$r_e$, demonstrating that dark matter makes
up an important part of the galaxy mass budget.

A recent optical study of the dark matter content of
three ellipticals by \citet{Romanowskyetal03} shows a quite different
result. All three galaxies have very low dark matter content, and two
are consistent with having no dark matter at all, at least out to
5$r_e$. At least one of these galaxies (NGC~4494) is X-ray faint, and two
of them are members of galaxy groups. We suggest that the X-ray luminosity
of early-type galaxies, which is dependent on the size of X-ray halo which
they can maintain, may be an indicator of their possession, or lack of, a dark
matter halo. This raises the question of how galaxies could lose (or gain)
an extensive dark matter halo, and how we might distinguish between
different processes which could affect such haloes. This is clearly a
question which deserves further consideration, and it seems likely that
both improved models and further observations of early-type galaxies will
be required before it can be answered.

\vspace{1cm}
\noindent{\textbf{Acknowledgments}\\
  We are grateful to S. Helsdon for the use of his 3-d gas properties
  software and J. Kempner for the use of his \chandra\ reduction software.
  We are also indebted to D. Forbes for his help in the early stages of the
  project, and to an anonymous referee for their efforts to improve the
  paper. This research has made use of the NASA/IPAC Extragalactic Database
  (NED) and Digitised Sky Survey (DSS). This research was supported in part
  by NASA grants NAG5-10071 and GO2-3186X.

\bibliographystyle{mn2e}
\bibliography{../paper}

\label{lastpage}

\end{document}